\documentclass[twocolumn]{aastex701}
\usepackage[utf8]{inputenc}
\usepackage{newunicodechar}
\newunicodechar{∼}{\textasciitilde}
\usepackage{hyperref}

\begin{document}

\title{The Binary-Binary Hierarchical System XY Leo: A Laboratory for Stellar Activity and Concealed Companions}

\author[orcid=0000-0002-3000-9829,sname='Koçak']{Dolunay Koçak}
\affiliation{Institute of Astronomy, The Observatories, Madingley Road, Cambridge CB3 OHA, UK}
\affiliation{Department of Astronomy and Space Sciences, Faculty of Science, Ege University, 35100, {\.I}zmir, Türkiye}
\email[show]{dolunay.kocak@google.com}  

\author[orcid=0000-0003-2380-9008,gname=Kadri,sname='Yakut']{Kadri Yakut} 
\affiliation{Department of Astronomy and Space Sciences, Faculty of Science, Ege University, 35100, {\.I}zmir, Türkiye}
\email[ky]{kadri.yakut@ege.edu.tr}

\author[orcid=0000-0002-1556-9449,gname=Chris,sname=Tout]{Chris Tout}
\affiliation{Institute of Astronomy, The Observatories, Madingley Road, Cambridge CB3 OHA, UK}
\email[cat]{cat@ast.cam.ac.uk }

\begin{abstract}
The hierarchical multiple  system XY~Leo, despite nearly 90 yr of observations, remains enigmatic. It offers a unique testbed for close binary evolution, involving processes like mass transfer, angular momentum loss, and the von Zeipel–Kozai-Lidov (ZKL) mechanism. Previously identified as a quadruple system, XY~Leo shows long-term orbital period modulations. Our new ground-based and Transiting Exoplanet Survey Satellite data suggest this may stem from either magnetic cycles or the influence of an unseen companion. While the latter remains speculative, both scenarios are discussed within a unified framework. Using all available photometric and spectroscopic data, we derived ultraprecise physical parameters for the contact binary XY Leo A as  M$_{\rm  A1}=0.629\pm0.009$\,M$_{\odot}$, M$_{\rm  A2}=0.865\pm0.012$\,M$_{\odot}$, R$_{\rm  A1}=0.739\pm0.007$\,R$_{\odot}$, R$_{ \rm A2}=0.855\pm0.008$\,R$_{\odot}$, L$_{\rm  A1}=0.271\pm0.026$\,L$_{\odot}$, L$_{\rm  A2}=0.288\pm0.030$\,L$_{\odot}$ and orbital separation a$_{\rm  A}=2.078\pm0.010$\,R$_{\odot}$ based on simultaneous solutions of light and radial velocity curves. The detached binary subsystem XY~Leo~B is confirmed to be on a wide $\sim$20-year orbit around the contact system. A second $\sim$23-year modulation is also detected, which may stem from either stellar magnetic activity or an additional, unseen companion. 
After removing both trends, a coherent residual modulation with a characteristic timescale of $14.2 \pm 0.8$~yr remains in the $O$--$C$ diagram, consistent with a magnetic activity cycle of Applegate type.
We modeled XY~Leo~A with the Cambridge \textsc{STARS (EV/TWIN)} code under non-conservative evolution, finding strong agreement between the tracks and observed parameters—highlighting the system’s value for testing multiple-star evolution.
\end{abstract}




\section{Introduction} 
\label{sec:int}

There are many different scenarios for the formation of contact systems. In all these scenarios, physical processes such as nuclear evolution, mass loss, and mass transfer play crucial roles in the transition from a detached binary configuration to a semi-detached and ultimately a contact system \citep{Yakut2005ApJ...629.1055Y}. If a close binary is part of a hierarchical multiple system—such as a triple or quadruple—the von Zeipel–Kozai–Lidov (ZKL) mechanism \citep{vonZeipel1910AN....183..345V,Kozai1962AJ.....67..591K,Lidov1962P&SS....9..719L} can significantly alter the orbital evolution and angular momentum distribution. Systems exhibiting both close binary evolution and outer dynamical perturbations serve as valuable laboratories to investigate such complex interactions. In this context, XY Leo is a particularly interesting system: previously identified as a quadruple, it shows long-term orbital period modulation. In this study, we explore whether this modulation arises from stellar magnetic activity or, alternatively, could indicate the gravitational influence of an additional unseen low-mass component.

The light variation of XY Leo (BD+18 2307), a short-period contact binary system of the W UMa type, was first discovered by \cite{Hoffmeister1934} and has since been extensively studied \citep{Koch1960AJ.....65..374K,Koch1978AJ.....83.1452K,Kaluzny1983AcA....33..277K,Hrivnak1985ApJ...290..696H,Pan1998Ap&SS.259..285P,Yakut2001IBVS.5042....1Y,Yakut2003A&A...401.1095Y,Bakis2005Ap&SS.296..131B}. The basic parameters of the system are listed in Table~\ref{Table:xyleo:basic:par}. XY Leo has been shown to exhibit strong indicators of magnetic activity, with an orbital period of 0.2841\,days \citep{Hilditch1981MNRAS.196..305H,Kocak2023PhDT,Kaluzny1983AcA....33..277K,Djuravsevic2006PASA...23..154D}. The Ca II H and K lines, indicative of chromospheric activity, were first detected in the system by \cite{Struve1959ApJ...130..137S}. Subsequent studies confirmed persistent Ca II H and K emission in the more massive component at all orbital phases \citep{Hrivnak1984ApJ...285..683H}. Many contact binary systems display asymmetry in their light curves—known as the O'Connell effect—where one maximum is brighter than the other, a phenomenon commonly attributed to starspots \citep{Yakut2005ApJ...629.1055Y}. Cyclical period variations are also widely observed in W UMa-type binaries and are often explained by magnetic activity cycles (e.g., \cite{Applegate1992}). For XY Leo, the asymmetry between the two maxima has been measured to be less than 0.04\,mag \citep{Yakut2009}. The difference in brightness between orbital phases 0.25 and 0.75, reported in earlier studies, is also apparent in the high-precision TESS light curves presented here.

Long-term orbital period variations in the XY Leo system have been investigated by several researchers \citep{Gehlich1972A&A....18..477G,Kaluzny1983AcA....33..277K,Hrivnak1985ApJ...290..696H,Pan1998Ap&SS.259..285P,Yakut2003A&A...401.1095Y,Kocak2023PhDT,Nasiroglu2025}. Early analyses suggested the presence of a third body orbiting the central contact binary, with an estimated orbital period of approximately 19.6 years \citep{Gehlich1972A&A....18..477G}. Based on the mass function derived from timing variations, the minimum mass of this object—later designated XY Leo B—was estimated to be 0.92\,M$_\odot$. Gehlich et al.\ also considered the possibility that this object might be a white dwarf, although this interpretation remains speculative. \cite{Yakut2009} attributed the combination of a sinusoidal variation and a long-term parabolic trend in the O–C diagram to the gravitational influence of a third component and to ongoing mass transfer between the inner binary stars. Assuming an inclination of $90^{\circ}$ for the outer orbit, \cite{Yakut2003A&A...401.1095Y} estimated the minimum total mass of XY Leo B to be $0.98 \pm 0.01$\,M$_\odot$, while \cite{Barden1987ApJ...317..333B} derived an orbital inclination of $29^{\circ}$ for the outer binary system formed by the third and fourth components.

The first radial velocity study of XY Leo was conducted by \cite{Struve1959ApJ...130..137S}, who reported semiamplitudes of $K_{A1} = 135$\,km\,s$^{-1}$ and $K_{A2} = 170$\,km\,s$^{-1}$, and a systemic velocity of $V_{\gamma,A} = -20$\,km\,s$^{-1}$. Later, \cite{Hrivnak1984ApJ...285..683H} obtained revised values of $K_{A1} = 108 \pm 2$\,km\,s$^{-1}$, $K_{A2} = 216 \pm 4$\,km\,s$^{-1}$, and $V_{\gamma,A} = -50 \pm 2$\,km\,s$^{-1}$. Spectral analysis by \cite{Barden1987ApJ...317..333B} revealed signatures associated with additional components, allowing the measurement of radial velocities for what is now known as the detached binary XY Leo~B. His analysis indicated that XY Leo is composed of two close binaries, forming a hierarchical quadruple system. He measured $K_{A1} = 124.1 \pm 2.8$\,km\,s$^{-1}$, $K_{A2} = 204.7 \pm 2.5$\,km\,s$^{-1}$, and $V_{\gamma,A} = -51.8 \pm 2.2$\,km\,s$^{-1}$ for the inner pair, and $K_{B1} = 46.1 \pm 0.7$\,km\,s$^{-1}$ and $K_{B2} = 64.6 \pm 1.8$\,km\,s$^{-1}$ for the outer binary. Subsequent high-resolution spectroscopic studies by \cite{Pribulla2007AJ....133.1977P}, \cite{Rucinski2007AJ....134.2353R}, and \cite{Pribulla2009AJ....137.3655P} confirmed these results and reported $K_{A1} = 144.7 \pm 1.1$\,km\,s$^{-1}$, $K_{A2} = 198.4 \pm 1.1$\,km\,s$^{-1}$, and $V_{\gamma,A} = -51.2 \pm 0.6$\,km\,s$^{-1}$. Unlike the earlier study, they detected only a single-lined spectrum for XY Leo B and estimated its light contribution as $L_{\rm B}/L_{\rm total} = 0.13$. Both groups classified the components of XY Leo B as mid-M type stars with an orbital period of 0.81\,days. \cite{Rucinski2007AJ....134.2353R} also estimated the orbital inclination of the outer binary to be $67^\circ$, and the projected separation between the A and B subsystems as $0\farcs061 \pm 0\farcs014$.

The physical parameters of the XY Leo~A contact binary have been determined in previous studies through simultaneous analysis of photometric and spectroscopic data. Reported component masses include 0.63 and 0.50\,M$_\odot$ \citep{Koch1978AJ.....83.1452K}, 0.87$\pm$0.04 and 0.44$\pm$0.02\,M$_\odot$ \citep{Hrivnak1985ApJ...290..696H}, and 0.82 and 0.50\,M$_\odot$ in both \cite{Yakut2006PhDT.......101Y} and \cite{Djuravsevic2006PASA...23..154D}. In this study, we compile and re-analyse all available data sets, including new ground-based observations and space-based photometry from TESS Sectors 45 and 46. The physical and orbital parameters of both binary systems (XY Leo~A and XY Leo~B) are derived using simultaneous modelling of light and radial velocity curves. TESS Sector 72 data are used exclusively for timing analysis of mid-eclipses.
The system shows a clear long-term modulation in its orbital period. While magnetic activity remains a strong candidate to explain this variation, we also explore the alternative possibility of a low-mass gravitational perturber in a wide orbit. Further observations will be needed to discriminate between these scenarios.
In Section~\ref{sec:obs} we describe the observations and data reduction procedures. 
In Section~\ref{sec:dataanalysis} we present the period change analysis, the modelling of the radial velocity curves, and the light-curve solution of the system. In Section~\ref{sec:Physical_par} we detail of the derivation of physical parameters for all four components. A discussion of the results and concluding remarks are given in the final section.

\begin{table}
\caption{Basic properties of XY Leo from the Simbad \citep{Wenger2000A&AS..143....9W}, Gaia \citep{gaia2021A&A...649A...1G} and TESS Input Catalogue (TIC) \citep{TESS2015JATIS...1a4003R}.}
\begin{tabular}{llll}
\hline
Identifying information &&&   \\
\hline
TIC                       &&& 358713471                    \\
2MASS ID                  &&& J10014043+1724328          \\
Gaia ID                   &&& 623076270045202048    \\
$\alpha_{2000}$           &&& 09:58:55.87             \\
$\delta_{2000}$           &&& +17:39:03.19             \\
\hline
Photometric  properties &&&   \\
\hline
G (Gaia)                  &&& $9\fm 293\pm0.009$      \\
B                         &&& $10\fm 66\pm0.06$             \\
V                         &&& $9\fm 68\pm0.03$            \\
R                         &&& 8\fm86           \\
2MASS J                   &&& $7\fm 691\pm0.023$           \\
2MASS H                   &&& $7\fm 149\pm0.036$            \\
2MASS K                   &&& $6\fm 990\pm0.034$            \\
\hline
Stellar properties &&&   \\
\hline
Spectral type             &&& K0 V            \\
Period (XY Leo A)      	  &&& 0.284\,d          \\
Period (XY Leo B)        &&& 0.805\,d          \\
$\pi_{\rm Gaia}$          &&& 14.9928 mas        \\
\hline
\end{tabular}
\label{Table:xyleo:basic:par}
\end{table}

\section{Observations} \label{sec:obs}

\begin{figure*}
\centering 
\includegraphics[width=0.75\textwidth,height=0.5\textheight,keepaspectratio]{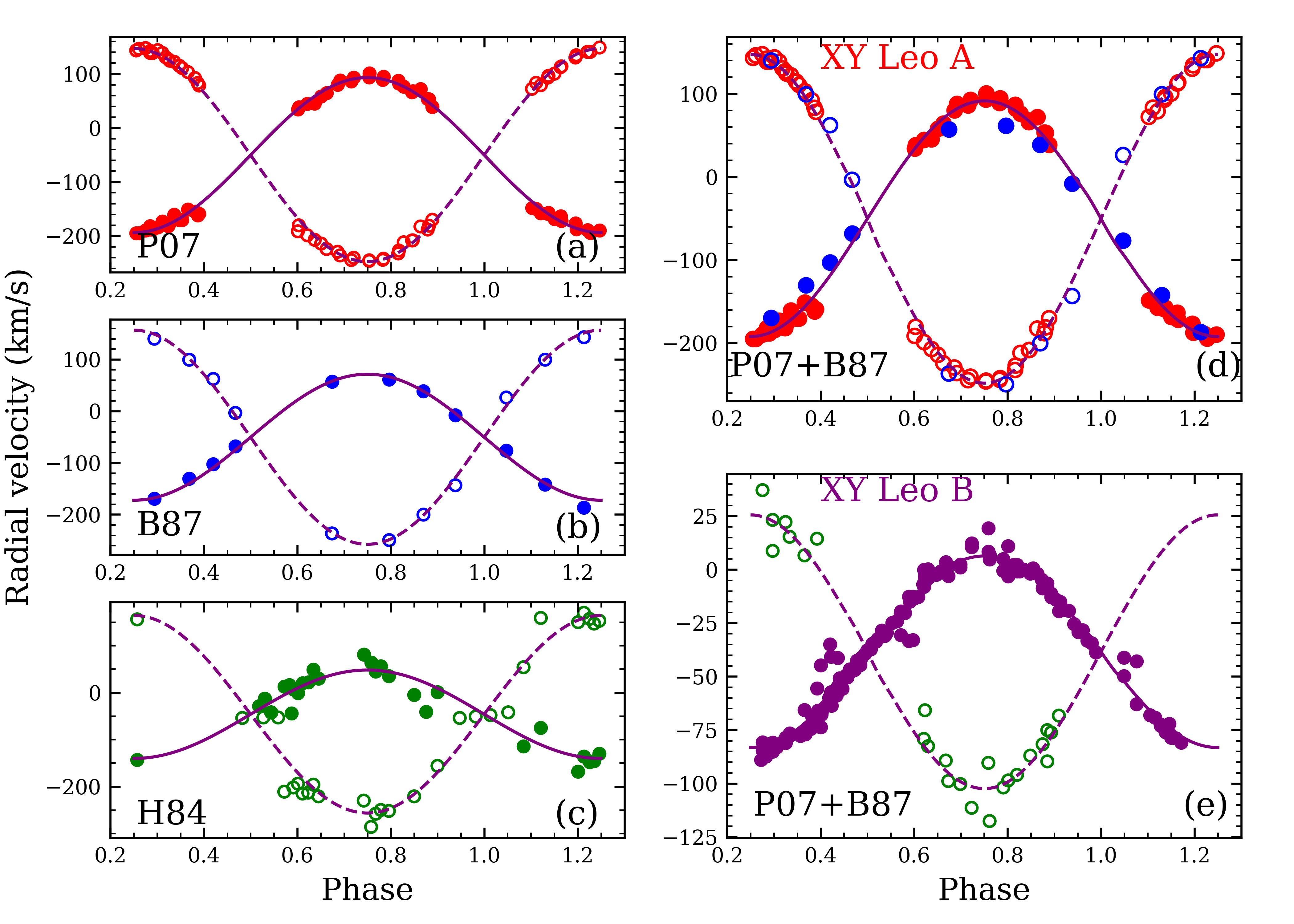}
\caption{Radial velocity observations and derived models of the contact binary system XY Leo A. The filled circular dots represent the primary component and the open circles the radial secondary. The solid line represents the primary star and the purple dashed line the secondary star. The re-analysed observations are taken from the studies published earlier \citep{Hrivnak1984ApJ...285..683H,Barden1987ApJ...317..333B,Pribulla2007AJ....133.1977P} (H84, B87, P07). See text for details.}
\label{Fig:xyleo_all_rvs}
\end{figure*}

\begin{figure}
\centering
\includegraphics[scale=0.85]{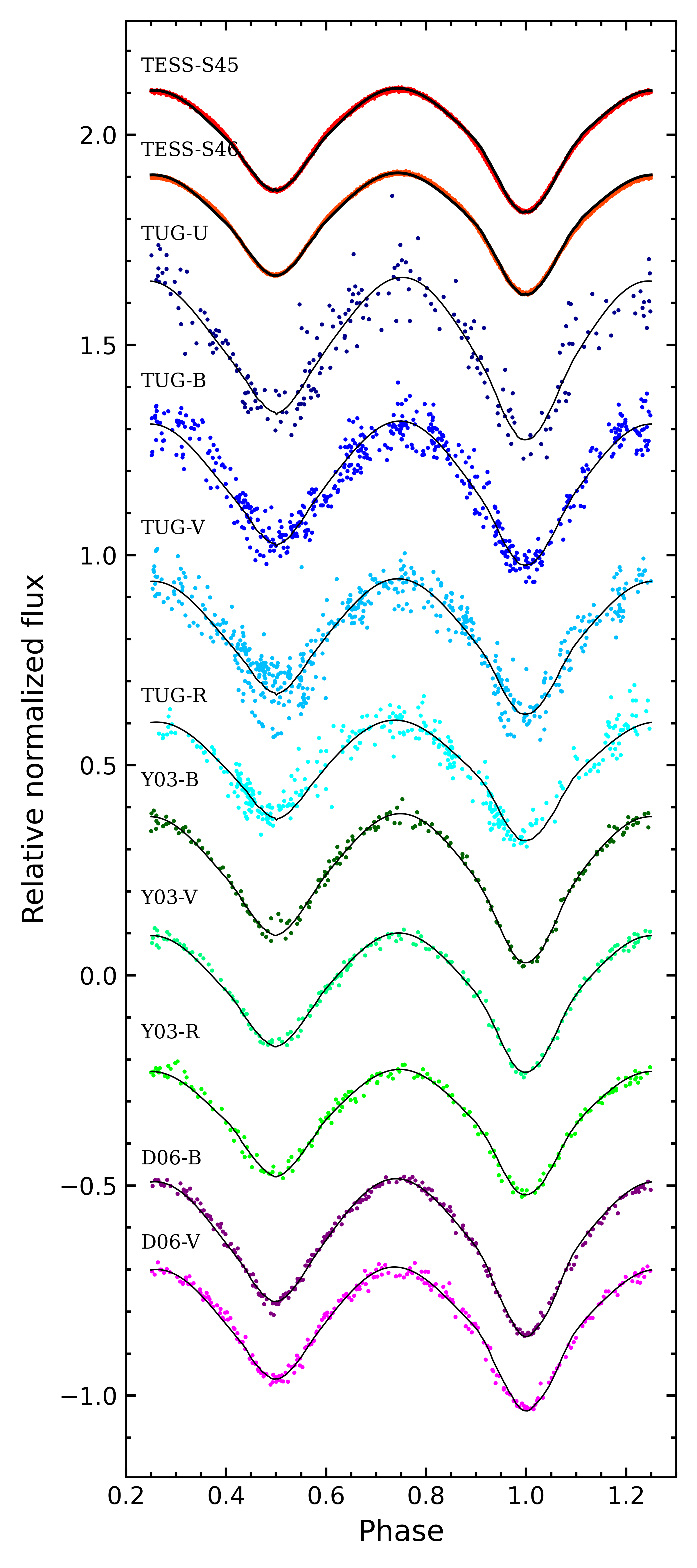}
\caption{Superposition of synthetic models (continuum, lines) derived from the simultaneous solution of the XY Leo A contact multiple system with multicolour ($UBVR$) newly obtained TUG T60, TESS Sector 45-46 and photometric observations available in $BVR$ filters published earlier \citep{Yakut2003A&A...401.1095Y,Djuravsevic2006PASA...23..154D}.}
\label{Fig:xyleo_all_lc}
\end{figure}

Since the discovery of its short-period photometric variability nearly 90 years ago, XY Leo has been observed extensively through both ground-based and space-based campaigns. We have analysed our new observations together with previously published data (Section~\ref{sec:int}), including photometric time series from the Transiting Exoplanet Survey Satellite (TESS) space telescope \citep{TESS2015JATIS...1a4003R}.
Spectroscopic observations of the system have been obtained by four different groups. Although the radial velocity measurements reported by Struve (1959) could not be included in this study due to the lack of published numerical data, the datasets from H84 \citep{Hrivnak1984ApJ...285..683H}, B87 \citep{Barden1987ApJ...317..333B}, and P07 \citep{Pribulla2007AJ....133.1977P} were used in our analysis. The combined radial velocity measurements used in this work are shown in Fig.~\ref{Fig:xyleo_all_rvs}. Similarly, the multicolour photometric light curves obtained by D06 \citep{Djuravsevic2006PASA...23..154D} and Y03 \citep{Yakut2003A&A...401.1095Y} were incorporated into our study. The available light curves were phase-folded using the orbital parameters derived from our analysis (see Section~3) and are shown in Fig.~\ref{Fig:xyleo_all_lc}.

Optical observations of XY Leo were conducted between 2018 and 2020 with the 60-cm robotic T60 telescope at the  TUG (TÜBİTAK National Observatory, Antalya) site under the Türkiye National Observatories. The observations were obtained with two CCD detectors, the iKon-L 936 BEX2-DD and the FLI ProLine 3041-UV cameras, operated on the TUG T60 telescope with OPTEC Bessell
Johnson--Cousins $UBVR$ photometric filters. Data reduction was performed with \textsc{AstroImageJ} \citep{Collins2017}, following standard procedures for bias and dark subtraction,
followed by flat-field correction. Aperture photometry was then performed to obtain differential light curves. The resulting multicolour UBVR photometry from TUG is listed in
Table~\ref{tab:xyleo:tugdata} for future reference.

In addition, the XY Leo system was observed by TESS during Sectors 45, 46, and 72, covering a total of 72 days.  Raw full-frame image (FFI) data were downloaded from the
MAST\footnote{\url{https://mast.stsci.edu}} archive
\citep{TESS_XYLeo_2025}. The TESS photometric time series were extracted and analysed
using the Lightkurve Python package \citep{Lightkurve2018ascl.soft12013L}, following the methodology adopted in our previous studies \citep{Yakut2015MNRAS.453.2937Y,
Cokluk2019MNRAS.488.4520C,Kocak2021ApJ...910..111K}. The combined light curves from TUG, the literature, and TESS are shown in Fig.~\ref{Fig:xyleo_all_lc}.

\begin{table}
\begin{center}
\caption{TUG--T60 multi-colour photometric measurements of XY~Leo obtained with 
Bessell $U$, $B$, $V$, and $R$ filters (Johnson--Cousins photometric system).}\label{tab:xyleo:tugdata}
\begin{tabular}{llll}
\hline
BJD (2400000+)	 & Phase	& Normalized Flux   & Filters\\
\hline
57139.630996	&	0.2619	&	1.033214	&	U	\\
57139.651428	&	0.2639	&	1.033458	&	U	\\
57139.671860	&	0.2659	&	1.033559	&	U	\\
57139.692292	&	0.2679	&	1.033646	&	U	\\
57139.712724	&	0.2699	&	1.033744	&	U	\\
57139.733156	&	0.2719	&	1.033801	&	U	\\
57139.753588	&	0.2739	&	1.033739	&	U	\\
57139.774020	&	0.2759	&	1.033825	&	U	\\
57139.794452	&	0.2779	&	1.034038	&	U	\\
57139.814884	&	0.2799	&	1.033900	&	U	\\
57139.835316	&	0.2819	&	1.034085	&	U	\\
57139.855749	&	0.2839	&	1.033757	&	U	\\
57139.876181	&	0.2860	&	1.033963	&	U	\\
57139.896613	&	0.2880	&	1.033953	&	U	\\
\hline
\end{tabular}
\end{center}
{\textbf{Notes.} The TUG--T60 observations listed here are shown together with other available light curves in Fig.~\ref{Fig:xyleo_all_lc}. 
The fourth column lists the photometric filter used for each observation ($U$, $B$, $V$, or $R$). 
Table \ref{tab:xyleo:tugdata} is published in its entirety in the electronic edition of the journal and on CDS only. 
A portion is shown here for guidance regarding its form and content.}
\end{table}

We obtained more than 240 light curves from the TESS observations. From these, we obtained a high number of minima. Instead of a large number of minima, we folded and read consecutive minima and calculated a total of 24 new minima times. The minima obtained are listed in Table~\ref{table:mintimes:xyleo}, together with all those found in the literature. The minima in Table~\ref{table:mintimes:xyleo} are distributed over a range of 68.9 years. This allows us to study the changes over such long periods. 


\begin{table}
\begin{center}
\caption{
A partial set of times of the minima as BJD (BJD $-$24\,00\,000). The reference codes (R) are as follows: (1) Minima times compiled from historical data and listed in the PhD thesis of D. Ko\c{c}ak \citep{Kocak2023PhDT}; (2) Published measurements from \citet{Nasiroglu2025}; (3) Newly extracted minima times from TESS observations. Table~\ref{table:mintimes:xyleo} is published in its entirety in the electronic edition of the journal and also available via the SIMBAD astronomical database (CDS).
}\label{table:mintimes:xyleo}
\begin{tabular}{llllll}
\hline
BJD	&	R	&	BJD	&	R	&	BJD	&	R\\ 
\hline
...&	...	&	... &	...	&	...	&	...	\\
...&	...	&	... &	...	&	...	&	...	\\
59527.12092	&	3	&	59542.31993	&	3	&	59559.93426	&	3	\\
59527.26217	&	3	&	59543.31455	&	3	&	59560.07604	&	3	\\
59529.96166	&	3	&	59545.01908	&	3	&	59568.03180	&	3	\\
59530.10299	&	3	&	59545.16074	&	3	&	59568.17353	&	3	\\
59532.80276	&	3	&	59554.11043	&	3	&	59570.87271	&	3	\\
59533.22840	&	3	&	59554.25204	&	3	&	59571.01460	&	3	\\
59542.17833	&	3	&	59556.95112	&	3	&	59573.71370	&	3	\\
59542.31935	&	3	&	59557.09321	&	3	&	59573.85519	&	3	\\
\hline
\end{tabular}
\end{center}
\end{table}

\section{DATA ANALYSIS} \label{sec:dataanalysis}

Detailed analyses of photometric and spectroscopic observations of an eclipsing binary system over many years can provide us with the parameters of the component stars and the orbital parameters of the system. In addition, the time variation of the mid-minima of the light curves obtained from photometric observations provides us with information about some of the processes affecting the orbit and thus the evolution of the binary system. These processes include physical processes such as mass transfer, mass loss, the existence of a third or fourth body orbiting the binary, and stellar activity. In this study, the period change analysis, radial velocity and light curve analysis techniques used for eclipsing binary systems are presented below.

\subsection{Period Change Analysis}\label{sec:periodchange}

The orbital period of a binary system may vary due to processes such as mass loss by stellar activity, mass transfer and the ZKL effect, which occurs in the presence of a third body. Processes that cause period changes in close and interacting systems can be detected by long-term observations. The follow-up observations of XY Leo have provided sufficient times of minima to understand the nature of the system. The existence of minimum observations of XY Leo over a relatively long period of time makes it an important candidate for probing such long-term astrophysical variations more reliably. For the TESS observations, we derived 24 new minima times using the method of \cite{Kwee1956BAN....12..327K}. These minima were obtained by folding consecutive light curves within individual TESS sectors and measuring the eclipse timings from the resulting combined light curves, which provide more reliable minima times. The newly obtained minima and all available minima are listed together in Table~\ref{table:mintimes:xyleo}.

\begin{figure}
\centering
\includegraphics[height=78mm,keepaspectratio]{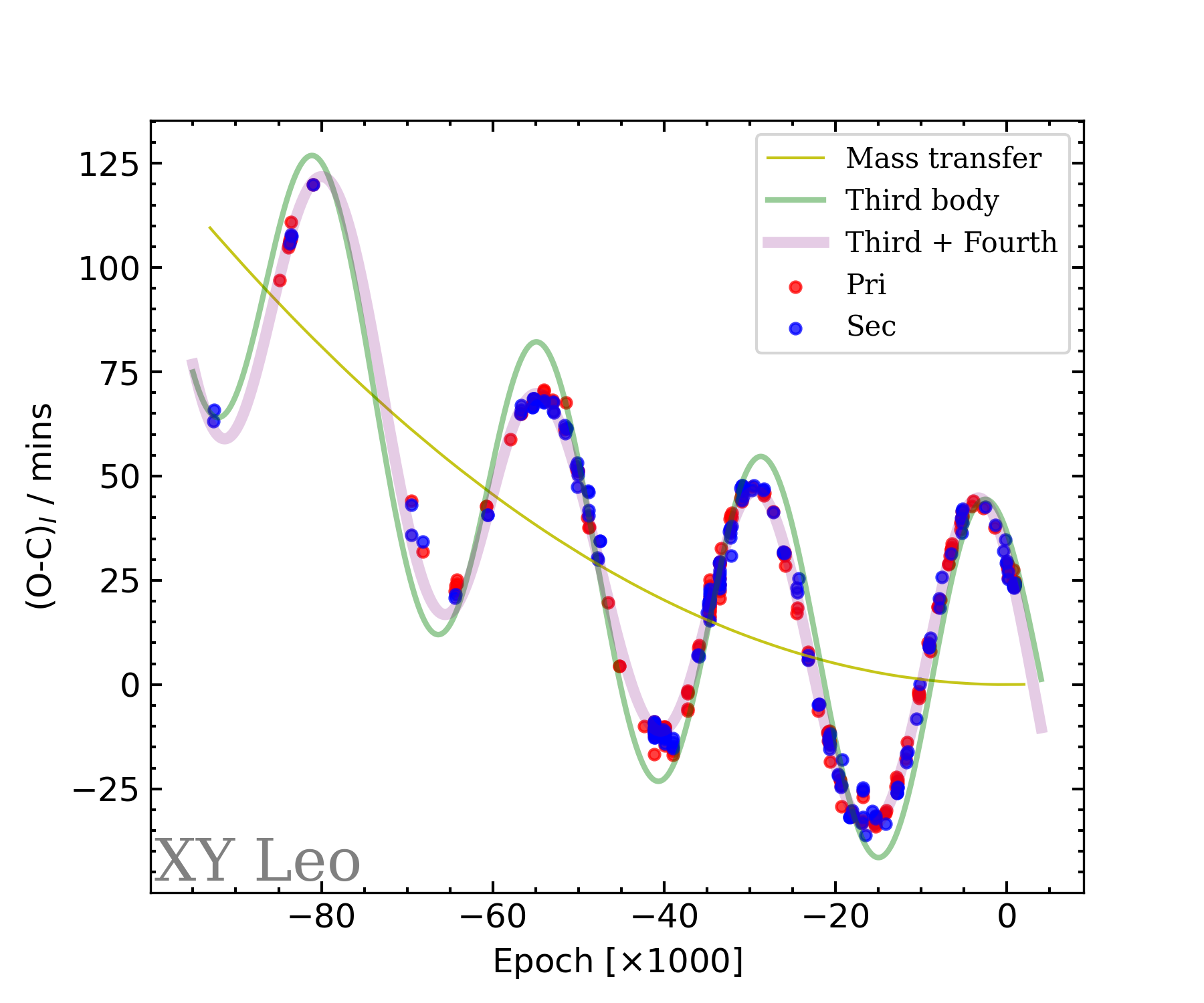}\\
\includegraphics[height=31mm]{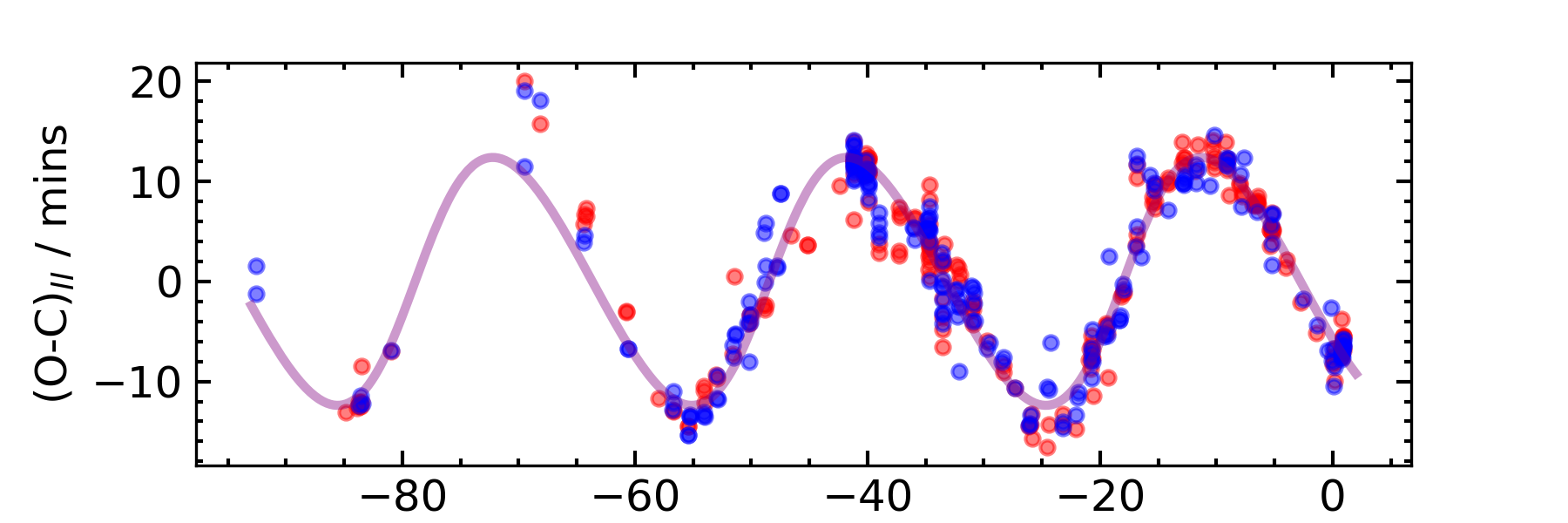}\\
\includegraphics[height=31mm]{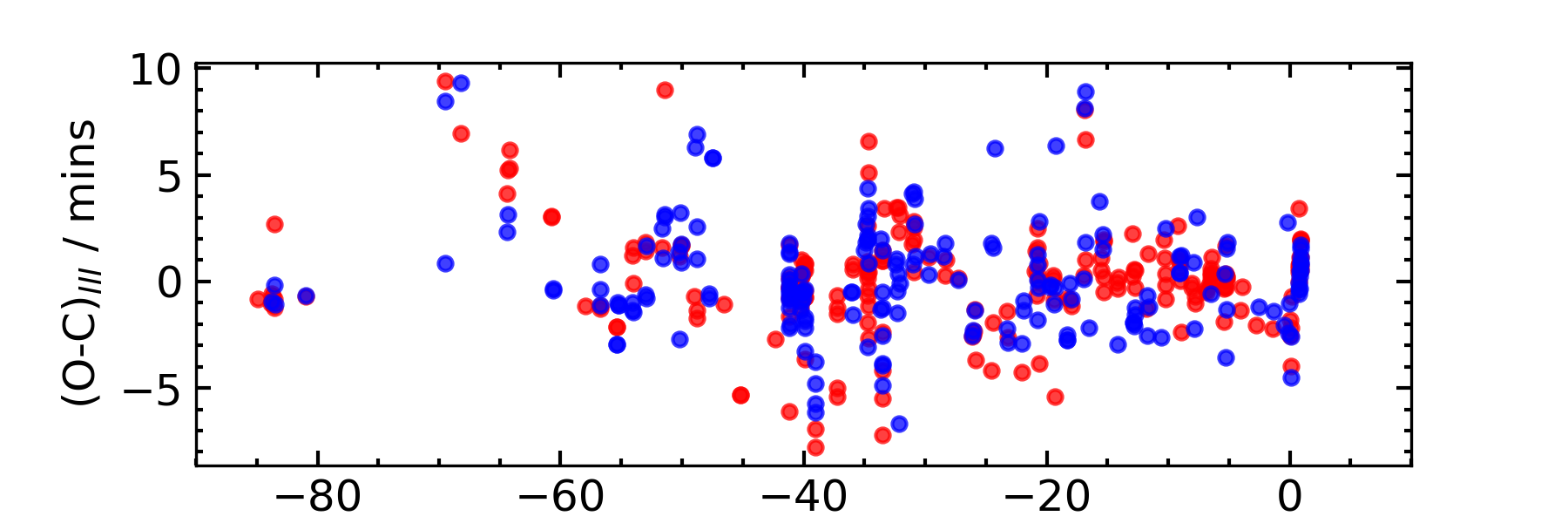}\\
\includegraphics[height=28.3mm]{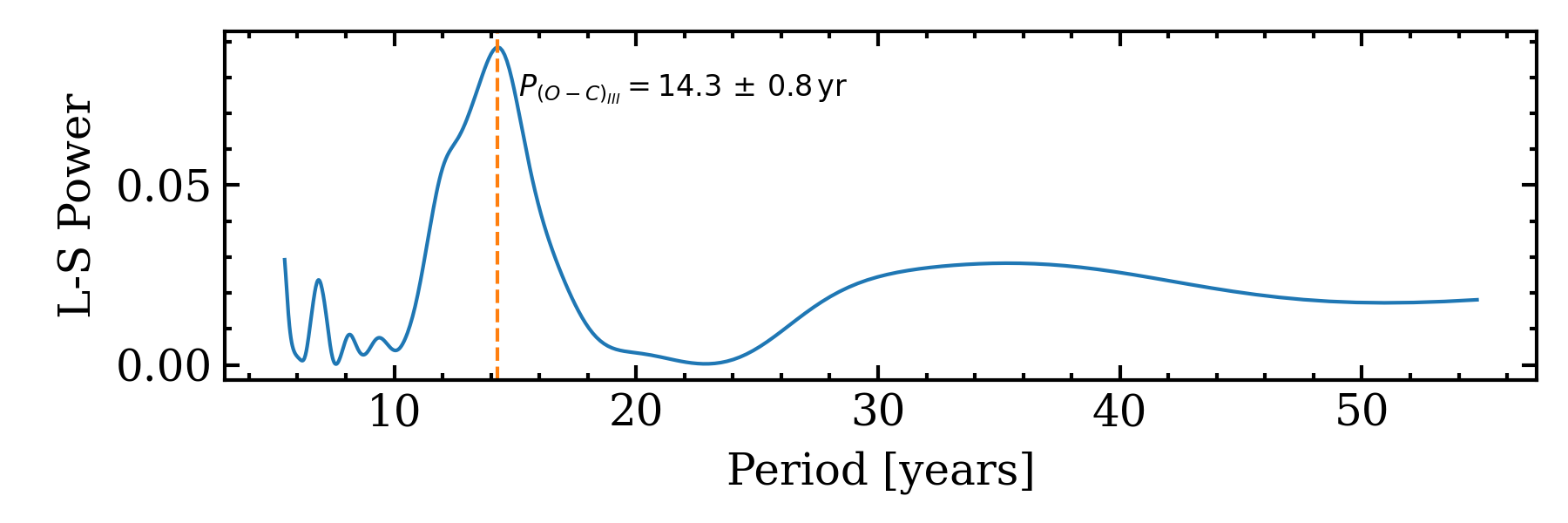}
\caption{Period change analysis of the XY~Leo~A system.
The panels display successive stages of the $O\!-\!C$ analysis.
The top panel shows the $O\!-\!C$ diagram fitted with two LITE terms and a long-term modulation.
The first LITE signal ($P \approx 20$~yr) is attributed to the well-established third body, XY~Leo~B.
The second LITE term (second panel, $P \approx 23$~yr) is included for completeness; however, magnetic activity is considered a more plausible explanation than an additional stellar component.
The third panel presents the final residuals, $(O\!-\!C)_{\mathrm{III}}$, which exhibit a coherent quasi-cyclic modulation.
The bottom panel shows the Lomb--Scargle periodogram of the $(O\!-\!C)_{\mathrm{III}}$ residuals, revealing a dominant cycle at $P_{(O-C)_{\mathrm{III}}}=14.2 \pm 0.8$~yr, consistent with a magnetic activity cycle of Applegate type.}
\label{Fig:xyleo_oc}
\end{figure}

\begin{figure}
\centering
\includegraphics[scale=0.8]{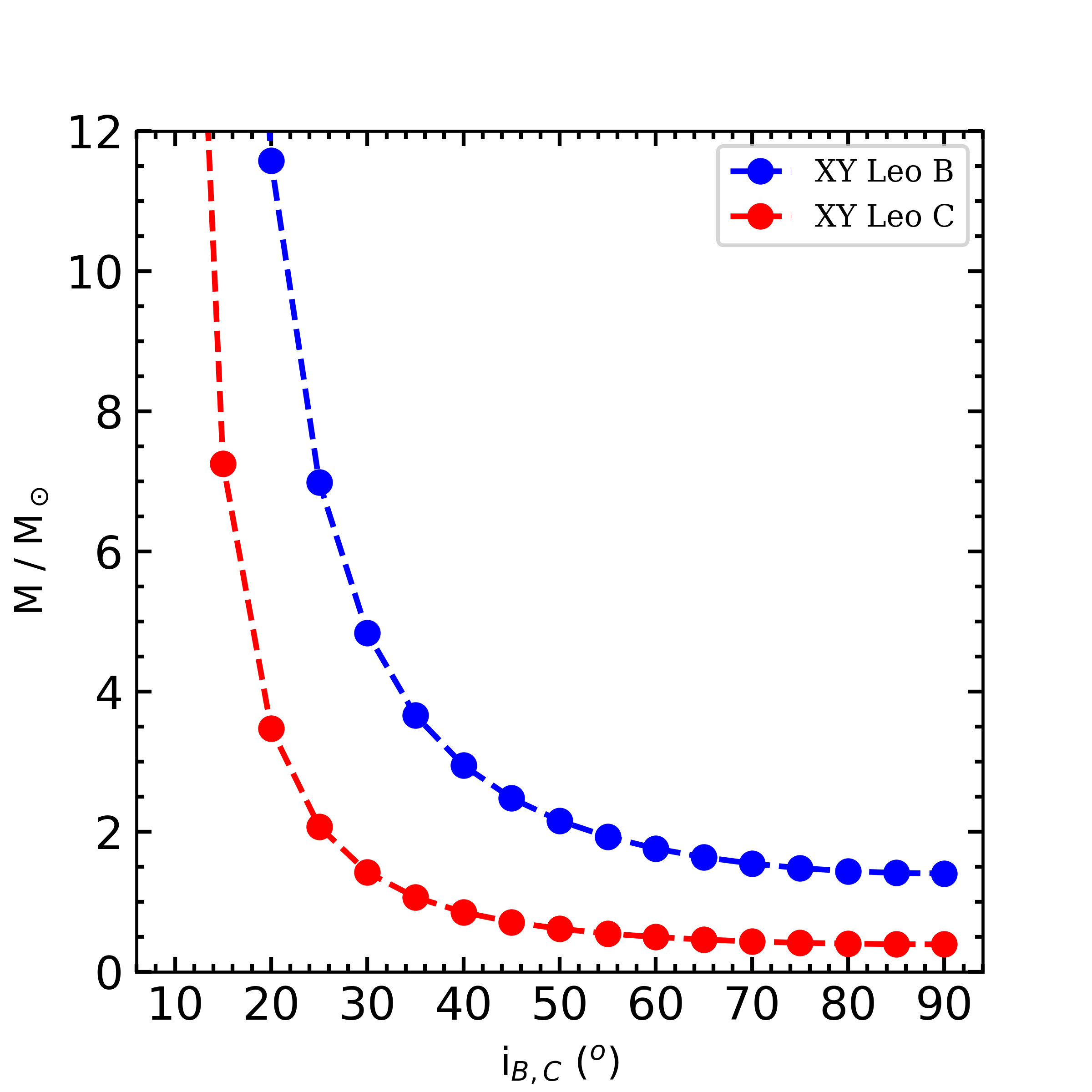}
\caption{Total mass estimates as a function of orbital inclination for the two LITE components. The first corresponds to the confirmed third body (XY~Leo~B), while the second signal (LITE2) is treated as speculative and may instead arise from magnetic activity. The curves assume Keplerian motion and are shown for illustrative purposes only; the reality of the second modulation remains uncertain due to potential dynamical instability.}
\label{Fig:xyleo:ocmass34}
\end{figure}

The overall O$-$C structure of XY~Leo is best described by four distinct components:
(i) a parabolic term caused by ongoing mass transfer within the contact binary, 
(ii) a well-established $\sim$20-year LITE signal produced by the detached binary subsystem XY~Leo~B, 
(iii) an additional $\sim$23-year modulation whose physical origin is ambiguous (either a dynamically problematic fifth component or an activity-related quasi-periodic variation), and 
(iv) a final low-amplitude $\sim$14-year residual modulation that is energetically consistent with the Applegate mechanism and most plausibly linked to magnetic cycles.

All obtained times of minima for the mid-eclipses in the light curve of the XY Leo system were analysed with a period change analysis program \citep{Zasche2009NewA...14..121Z} coded in {\tt MATLAB}. In the analysis, different weights were assigned to the minima according to the typical timing precision of the datasets. Minima derived from the \textit{TESS} light curves were given twice the weight of the ground-based measurements, reflecting the higher time resolution and photometric precision of the space-based observations. We previously applied a similar method in our study on CC Com \citep{Kocak_2024}.
The initial modelling assumed orbital period variations resulting from both mass transfer and the light travel time effect (LITE), caused by the known tertiary companion (XY Leo B). Overlapping solutions were explored according to Equations~\ref{Eq:xyleo:3rdbody} and \ref{Eq:xyleo:3rdbody2}, under the assumption  of sinusoidal and parabolic trends. The first two terms in Equation~\ref{Eq:xyleo:3rdbody} are linear, and the third is parabolic, describing secular period evolution owing to mass transfer. The $\tau$ term represents the LITE contribution from the motion of the eclipsing binary around a common barycentre with the outer companion, parameterised by the orbital elements of XY Leo B \citep{Irwin1952ApJ...116..211I}. To estimate the minimum mass of this third component, the expression in Equation~\ref{Eq:xyleo:fm} is used. The symbols $i^{\prime}$, $e^{\prime}$, $v^{\prime}$, and $\omega^{\prime}$ represent the orbital inclination, eccentricity, true anomaly, and argument of periastron of the third component, respectively.
\begin{equation}
\mathrm{Min\,I}=T_{0}+P_{0} E+\frac{1}{2} \frac{d P}{d E} E^{2} + \tau  
\label{Eq:xyleo:3rdbody}
\end{equation}
\begin{equation}
 \tau = \frac{a_{12} \sin i^{\prime}}{c}\times\left[\frac{1-e^{2}}{1+e^{\prime} \cos v^{\prime}} \sin \left(v^{\prime}+\omega^{\prime}\right)+e^{\prime} \sin \omega^{\prime}\right]
\label{Eq:xyleo:3rdbody2} 
\end{equation}

\begin{equation}
f(m)=4 \pi^2 \frac{\left(a_{12} \sin i\right)^3}{G P_{m}^2}=\frac{\left(M_3 \sin i\right)^3}{\left(M_1+M_2+M_3\right)^2},
\label{Eq:xyleo:fm}
\end{equation}

We analysed the orbital period variations of the system using all available minima listed in Table~\ref{table:mintimes:xyleo}. As shown in Fig.~\ref{Fig:xyleo_oc}, both sinusoidal and parabolic trends are apparent. We initially interpreted the parabolic variation as evidence of mass transfer within the contact system XY Leo A, while the long-term sinusoidal trend was modelled as a light-travel time effect owing to the orbit of XY Leo A around a common barycentre with the detached binary XY Leo B. However, when examining the residuals $(O-C)_{II}$, a remaining modulation is still noticeable. To explore this, we considered additional explanations, including the possibility of magnetic activity or a further low-amplitude periodicity that may suggest additional dynamical complexity.

The results obtained from modelling with a third and an additional hypothetical component are given in Table~\ref{tab:xyleo:OC}, together with their associated uncertainties. The period change analysis reveals a $20.17 \pm 0.12$\,yr signal consistent with the gravitational influence of the known third body. An additional $\sim$23\,yr modulation is present in the residuals but its origin remains uncertain. Table~\ref{tab:xyleo:OC} also provides estimated masses for different assumed orbital inclinations. We also find evidence of mass transfer within the contact system. The quadratic term in the O--C diagram indicates a secular increase 
of the orbital period of approximately 0.21 seconds per century. 
The inferred mass transfer rate is $6.7 \times 10^{-8}$\,M$_\odot$\,yr$^{-1}$, 
consistent with mass transfer from the less massive to the more massive component.

The top panel of Fig.~\ref{Fig:xyleo_oc} shows the effects of mass transfer (purple parabolic curve) and the known third body, XY~Leo~B (blue dashed line), along with a combined fit that includes an additional sinusoidal modulation (red solid line).
The dashed parabola represents the secular period increase attributed to mass transfer in the XY~Leo~A contact binary. The blue sinusoid corresponds to the $\sim$20-year variation associated with XY~Leo~B.
The red curve represents a two-component fit; however, we emphasise that the physical origin of the second, longer-term modulation remains unclear.
This signal may reflect a low-amplitude gravitational effect or be the result of magnetic activity.
The second panel isolates the longer-term modulation after subtracting the parabolic trend.
Although the modulation can be approximately modelled with a second sinusoid, its origin remains uncertain.
It could reflect a magnetic cycle rather than Keplerian motion, particularly because the observed variability does not cover a full, well-sampled cycle.
Unlike strictly periodic variations caused by orbiting bodies, magnetic activity is expected to produce quasi-cyclic, non-sinusoidal patterns, as observed in the Sun.
Residuals ($(O\!-\!C)_{\mathrm{III}}$) shown in the third panel reveal a remaining low-amplitude variation with a peak-to-peak amplitude of about 12 minutes.
A Lomb--Scargle analysis of these final residuals yields a dominant quasi-cyclic timescale of $14.2 \pm 0.8$~yr, supporting an interpretation in terms of magnetic activity rather than an additional Keplerian companion. Continued monitoring will be necessary to determine whether this modulation repeats in a coherent and dynamically meaningful way.

Period change analysis, when applied to long-term data sets, can provide insights into the presence of additional bodies orbiting a long-period binary system. However, distinguishing between different mechanisms influencing period variations remains challenging.
The interpretation of the cause of the variation in the residuals from the O-C analysis of most binary systems depends on whether it is owing to a third body or activity. Many similar studies of period variations in close binaries have been reported in the published record. However, these interpretations are generally influenced by the fact that stellar activity is cyclic rather than strictly periodic. Although not a general expectation, due to the nature of the activity, the residuals are often interpreted as showing cyclic variations in some systems, while others exhibit more regular sinusoidal-like patterns. However, distinguishing between these categories is not always straightforward, and further statistical analysis and long-term monitoring would be required to better classify these variations.

The interpretation of the 23-year residual variation as originating from an additional orbiting body beyond the known components remains highly speculative. Given that the known 20-year outer orbit (attributed to XY~Leo~B) already constrains the system’s dynamical architecture, the introduction of another companion with a nearby 23-year orbital period would imply a period ratio of only 1.2. According to hierarchical stability criteria \citep[e.g.][]{Fabrycky2007ApJ...665..754F,Mardling2001MNRAS.321..398M,Tokovinin2008MNRAS.389..925T}, such a configuration would likely be dynamically unstable unless it includes very specific conditions such as a high mutual inclination or an extremely low companion mass. As a result, we refrain from designating this signal as a distinct stellar component and instead consider it more plausibly as a manifestation of stellar magnetic activity, analogous to the cyclic variations observed in solar-type and W~UMa-type stars. Nonetheless, further studies—including dedicated $N$-body simulations and high-resolution spectroscopic monitoring—are needed to investigate the origin and stability of this signal.

Given the spectroscopic evidence of strong H$_\alpha$ and Ca II H \& K emission from the components of XY Leo~B reported by \citet{Barden1987ApJ...317..333B}, and its short orbital period of 0.8 days suggestive of RS CVn-type magnetic activity, it is plausible that the residual $\sim$14-year modulation observed in the $O-C$ diagram (Bottom panel of \ref{Fig:xyleo_oc}) originates from magnetic cycles within the XY Leo~B system itself, rather than the inner contact binary.

The interpretation of long-term cyclic variations in $O-C$ diagrams of close binaries has long been debated in the literature. In systems with cool and magnetically active components, such modulations are often associated with magnetic activity cycles \citep{Hall1989, Tout1991, Applegate1992}. Changes in the internal angular momentum distribution and magnetic field geometry within the convective envelope of the active star can lead to variations in the stellar quadrupole moment, which in turn modulate the orbital period \citep{Lanza1998}. The relationship between these orbital period modulations and stellar spot cycles, although not always one-to-one, appears to be influenced by the star’s rotation rate and internal dynamo properties \citep{Lanza1999, Donati2003}. These effects are particularly relevant in contact or semi-detached binaries, where synchronous rotation enhances magnetic activity and may trigger observable period modulations even in the absence of a physical third body. We therefore explore the possibility that the observed 23-year modulation may be of magnetic origin, and assess its energetic feasibility in Section~5.

\subsection{Modeling Radial Velocity Curves}

Four different groups have published radial velocity measurements of XY Leo system. First, in 1959, \cite{Struve1959ApJ...130..137S} obtained the double-lined radial velocity parameters of the system. Based on the Ca II H and K lines \cite{Hrivnak1984ApJ...285..683H} obtained radial velocity parameters of the contact binary XY Leo A. \cite{Barden1987ApJ...317..333B} was able to obtain the radial velocities of both the contact binary system XY Leo~A and the detached binary system XY Leo~B separately and obtained their parameters. Most recently, \cite{Pribulla2007AJ....133.1977P} obtained the radial velocity of the contact binary system as well as the single-lined radial velocity of the detached binary system XY Leo~B. Each group quoted different radial velocity parameters. This is because the accuracy of the observations differs because of the different instruments used as well as the techniques used to find the parameters. Here, we reanalyzed all previously published radial velocity data using modern techniques to refine the system’s orbital parameters. Because the data were not published, we could not use the observations made by \cite{Struve1959ApJ...130..137S} in our analyses. 

The observations obtained by \cite{Hrivnak1984ApJ...285..683H}, \cite{Barden1987ApJ...317..333B} and \cite{Pribulla2007AJ....133.1977P} were solved separately and the data obtained by \cite{Barden1987ApJ...317..333B} and \cite{Pribulla2007AJ....133.1977P} were combined (P07+B87). The dataset of \cite{Hrivnak1984ApJ...285..683H} was analysed separately because it does not contain radial velocity measurements for XY~Leo~B, and its inclusion in the combined solution does not improve the orbital constraints. In contrast, the datasets of \cite{Barden1987ApJ...317..333B} and \cite{Pribulla2007AJ....133.1977P} provide radial velocities for both components and have significantly higher observational precision, allowing a more reliable combined orbital solution. The data used in the solution and the synthetic models of the radial velocities obtained are given in Fig.~\ref{Fig:xyleo_all_rvs} and Table~\ref{tab:XYLeo:RVs} with the resulting parameter errors. During the combined analysis of the datasets P07 and B87, we gave twice as much weight to the datasets of P07 because of their higher observational precision.

For the radial velocity analysis, we employed the same Monte Carlo-based fitting procedure that we previously used in the studies of multiple binary systems in the open clusters NGC~188 and NGC~2506 (see \citet{Yakut2025a,Yakut2025b}). Specifically, we used a custom Python code that performs a robust fitting of Keplerian orbital parameters using synthetic datasets generated from observational uncertainties. For each system, we generated $10^4$ realizations of the RV data and fitted orbital solutions using a least-squares optimization scheme. The distribution of resulting parameters allows us to estimate median values and $1\sigma$ uncertainties.In cases where multiple orbital solutions produce statistically indistinguishable residuals (i.e., $\Delta\chi^{2} < 1$), the combined set of these viable fits is used to ensure a consistent and statistically meaningful determination of the orbital parameters.

In the previous section, we used a different technique, period change analysis, which suggests that the XY Leo system could be a quintuple system. To avoid confusion, it should be noted again that XY Leo~A consists of a contact binary system with a period of 0.284 d in the inner orbit, while XY Leo~B consists of a detached close binary system with a period of 0.805 days orbiting the inner system with a period of 19.9\,yr. Additional, period variation analysis indicates the possible presence of an outer component with an orbital period of 23\,yr, though no radial velocity measurements have been detected to confirm this. Alternative explanations, such as stellar activity, may also contribute to the observed period variations. The results of the radial velocity analyses are in Figs~\ref{Fig:xyleo_all_rvs}a-d for XY Leo~A and Fig~\ref{Fig:xyleo_all_rvs}e for the XY Leo~B system with observations (open and filled circles). The results obtained are given in Table~\ref{tab:XYLeo:RVs}, together with the errors.

\begin{table}
\caption{Parameters related to the well-known 19.7-year orbit (LITE) and possible 20-year and 23-year orbits (LITE2) of the contact binary system XY~Leo A.} 
\begin{tabular}{llll}
\hline
Parameter                     & LITE              & LITE2          & Unit          \\
\hline $T_o$ (BJD)            & 59309.4843(5)     & 59309.4804(22) & day           \\
$P_{\rm orb, A12}$            & 0.2841040(1)      & 0.2841039(1)   & day           \\
                              &                   &                &               \\
$P_{\rm orb, B}$              & 19.7(2)           & 20.17(11)      & yr            \\
$T_B$  (BJD)                  & 58750(331)        & 56841(492)     & day           \\
$e_B$                         & 0.0253(4)         & 0.027(17)      &               \\
$f(m_B)$                      & 0.2174(2)         & 0.468(5)       & M$_{\odot}$   \\
$m_{B;i'=30^\circ}$           & 3.37(6)           & 5.7(1)         & M$_{\odot}$   \\
$m_{B;i'=60^\circ}$           & 1.33(3)           & 1.9932)        & M$_{\odot}$   \\
$m_{B;i'=90^\circ}$           & 1.08(2)           & 1.58(2)        & M$_{\odot}$   \\
$P_{\rm orb, C}$              &                   & 8318(214)      & day            \\
$P_{\rm orb, C}$              &                   & 22.8(6)        & yr             \\
$T_{\rm C}$ (HJD)             &                   & 54029(197)     & day            \\
$e_{\rm C}$                   &                   & 0.20(5)        &                \\
$f(m_C)$                      &                   & 0.0135(2)      & M$_{\odot}$    \\
$m_{C;i'=30^\circ}$           &                   & 2.07(5)        & M$_{\odot}$    \\
$m_{C;i'=60^\circ}$           &                   & 0.69(3)        & M$_{\odot}$    \\
$m_{C;i'=90^\circ}$           &                   & 0.54(2)        & M$_{\odot}$    \\
\hline
\end{tabular}
\label{tab:xyleo:OC}
\end{table}

\begin{table*}
\begin{center}
\caption{Radial velocities solutions of contact binary system XY Leo A and detached binary system XY Leo B. The standard errors $\sigma$ are given in parentheses in the last digit quoted, $T_{0}$ is given as ${\rm HJD}/{\rm{d} - 2453812}$.}
\label{tab:XYLeo:RVs}
\begin{tabular}{llllllll}
\hline
 &&\textbf{XY Leo A}&&&&\textbf{XY Leo B}&\\
\hline
                                    &	H84	     &	B87	      &	P07	     &	HBP	      &	BP	        &                                        &            \\
\hline
$T_{0, A}$                            &	0.180	 &	0.189     &	0.195	 &	0.195	  &	0.1949(4)	&$T_{0, B}$                               &	70.1277(28) \\
$P_{\rm orb,A12}/{\rm d}$ 	          &	0.284101 &	0.284103  &	0.284103 &	0.284103  &	0.284103(1)	&$P_{\rm orb,B12}/{\rm d}$ 	              &	0.804726(1) \\
$V_\gamma, {\rm A12}/{\rm km\,s^{-1}}$& -46.1	 &	-50.6	  &	-50.5	 &	-51.4	  &	-50.5(7)    &$V_{\gamma, {\rm B12} }/{\rm km\,s^{-1}}$&	-38.5(5)    \\
$K_{\rm A1}/{\rm km\,s^{-1}}$         &	210.4	 &	205.9	  &	197.1	 &	194.8	  &	197.8	    &$K_{\rm B1}/{\rm km\,s^{-1}}$            &	45.0(1.3)   \\
$K_{\rm A2}/{\rm km\,s^{-1}}$         &	94.2	 &	121.3	  &	143.4	 &	135.0	  &	141.6	    &$K_{\rm B2}/{\rm km\,s^{-1}}$            &	64.3(1.9)	\\
$q_{\rm A} = m_{\rm A1}/m_{\rm A2}$   &	0.448	 &	0.589	  &	0.7275	 &	0.693	  &	0.716(7)    &$q_B = m_{\rm B2}/m_{\rm B1}$   	      &	0.699(22)   \\
a$_{\rm A1}$sin$i/R_{_\odot}$         &	1.181	 &	1.156	  &	1.107	 &	1.093	  &	1.110	    &a$_{\rm B1}$sin$i/R_{_\odot}$            &	0.715	    \\
a$_{\rm A2}$sin$i/R_{_\odot}$         &	0.529	 &	0.681	  &	0.805	 &	0.758	  &	0.795       &a$_{\rm B2}$sin$i/R_{_\odot}$            &	1.022	    \\
m$_{\rm A1}$sin$^3i/M_{_\odot}$       &	0.257	 &	0.382	  &	0.490	 &	0.432	  &	0.480	    &m$_{\rm B1}$sin$^3i/M_{_\odot}$          &	0.064	    \\
m$_{\rm A2}$sin$^3i/M_{_\odot}$       &	0.575	 &	0.649	  &	0.673	 &	0.624	  &	0.671       &m$_{\rm B2}$sin$^3i/M_{_\odot}$          &	0.045	    \\
\hline
\end{tabular}
\end{center}
\end{table*}

\subsection{Light curve modelling of ground- and space based photometric data sets}

In addition to the newly obtained four-colour TUG photometric observations, we also used TESS data sets, the solution of which has not been used in any previous study. The previously published light curves obtained by \cite{Yakut2003A&A...401.1095Y} (Y03) and \cite{Djuravsevic2006PASA...23..154D} (D06) were also used in our analyses. In total, eleven light curves were analysed in five different filters. Unlike monochrome observations, light curve analysis with multi-colour observations gives us more reliable information about atmospheric parameters, especially surface temperatures. In addition, observations from highly sensitive photometric instruments, such as the TESS, offer an opportunity to determine geometric parameters with great precision. Therefore, analyses with observations obtained with different colours and sensitive instruments are important to obtain precise and more reliable parameters. 

In the previous subsection, different solutions of the radial velocities were given and the joint solution of P07 and B87 gave more reliable results. In this section, the radial velocities of P07 and B87 and all available light curves are solved altogether. The Phoebe and Wilson-Devinney programmes \citep{Wilson1971ApJ...166..605W,Wilson1979ApJ...234.1054W,prsa2005ApJ...628..426P} were used to fit the TESS data and the radial velocity curves simultaneously to obtain the orbital and physical parameters of the system.

The mean temperature of the cooler star T$_{\rm A1}$=4850 (170)~K according to the colour index, the logarithmic limb darkening coefficients (x$_{\rm A1}$=x$_{\rm A2}$) \citep{Claret2018A&A...618A..20C}, the albedos (A$_{\rm A1}$=A$_{\rm A2}$) \citep{Rucinski1969AcA....19..245R} and the gravitational darkening coefficients (g$_{\rm A1}$=g$_{\rm A2}$) \citep{Lucy1967ZA.....65...89L} are taken as fixed parameters during the analysis. The adjustable parameters are $a$, the semi-major axis of the relative orbit, $i$ the orbital inclination, V$_\gamma$ the radial velocity of the binary centre of mass, T$_{\rm 0}$ the epoch of primary minimum, P$_{\rm orb}$ orbital period of inner system, T$_{\rm A2}$, the spot parameters, $q$ the mass ratio, $\Omega _{\rm A1,A2}$ the potential of the components, the luminosities and the third light contribution (l$_{\rm B}$). 

To ensure a robust and global solution, a large number of iterations were performed, with all relevant parameters left free simultaneously. The analysis was halted only when the final corrections fell well below the estimated uncertainties, indicating convergence to an optimal set of parameters.
The parameters obtained are given in Table~\ref{table:xyleo:lcsol} together with their uncertainties. The agreement of the synthetic light curve models obtained with these results with the observations is also shown in Fig.~\ref{Fig:xyleo_all_lc}. It is seen that the models obtained with all data sets together are in very good agreement with the observations. Another agreement with our solutions is supported by the spectroscopic study conducted by \cite{Pribulla2009AJ....137.3655P}. They estimated, from the triple-Gaussian fits, that the broadening function around the quadratures of the contact binary is about L$_{\rm B1+B2}$/L$_{\rm A1+A2}\approx 0.13$. This is very similar to our results obtained by photometric light curve modelling. The Roche lobe filling factor of the system is calculated with the equation $f=(\Omega-\Omega_{\rm i})/(\Omega_{\rm o}-\Omega_{\rm i})$, where $\Omega_{\rm i}$ and $\Omega_{\rm i}$ are the dimensionless inner and outer potentials and $\Omega$ is the dimensionless potential on the common surface. The light curve analysis shows that the filling factor of the contact system XY Leo~A is 6.3 per cent.

From the light curves in Fig.~\ref{Fig:xyleo_all_lc} and the TESS observations, it can be seen that the system exhibits changes in the maxima of the light curves. These variations are usually attributed to the presence of spots on the stellar surface, also known as the O'Connell effect, which is the result of a change in the brightness of the star's surface. 
Although several modern light curve modelling tools (e.g., \texttt{PHOEBE2} \citep{prsa2005ApJ...628..426P}, \texttt{starry} \citep{Luger2019}, \texttt{fleck} \citep{Morris2020}) are capable of modeling spotted stellar surfaces with a wide range of parametric configurations, it is well known that such spot models often suffer from significant degeneracy and are rarely unique. Spot latitude, size, and contrast may trade off to produce equally acceptable fits, making the physical interpretation non-trivial. Therefore, while we attempted spotted model fits to the TESS and ground-based light curves, we do not claim a unique or physically constrained spot configuration.

Using the high-precision TESS data, we find that the starspots are on average about 10 per cent cooler than the stellar photosphere and occupy a region covering approximately 3.6 per cent of the total stellar surface.
These ratios vary slightly in many light curves. It is not possible to determine the coordinates and properties of the spots, because of lack of the uniqueness of the problem, but we point out that the system is active and this activity has an effect on the light variations and period variations.

\begin{table}
\caption{Light curve analysis results and their formal
1$\sigma$ errors.  See text for details.}\label{table:xyleo:lcsol}
\begin{tabular}{ll}
\hline
$T_{0}$ as ${\rm JD}/{\rm{d} - 24~00000}$   		& 59527.11987(9) \\
$P_{\rm orb}/{\rm d}$                               		& 0.2840981(7)   \\
Orbital inclination, $i$ ($^\circ$)             & 66.5(2)	 \\
$\Omega _{\rm A1,A2}$                             & 4.301(27)	  \\
$q = m_{\rm A2}/m_{\rm A1}$                       & 1.375    \\
$T_{\rm A1}/{\rm K}$                        & 4850 (170)      \\
$T_{\rm A2}/{\rm K}$                        & 4578 (160)   \\
Luminosity ratios:                       	& 				 \\
$l_{\rm A}/l_\textrm{A1+A2}$                & 47.3\%         \\
$l_{\rm B}/l_\textrm{A+B+C}$                & 13.3\%         \\
The fractional radius                       &               \\
$r_1 = R_{\rm A1}/a$                        & 0.35542(5)     \\
$r_2 = R_{\rm A2}/a$                        & 0.41147(5)     \\
Filling factor, f(\%)                        & 6.3             \\
\hline
\end{tabular}
\end{table}

\section{PHYSICAL PARAMETERS OF THE SYSTEM} \label{sec:Physical_par}
Spectral and photometric observations of binaries or multiple stars allow the precise determination of the orbital elements of these systems, as well as the absolute dimensions of their components, such as temperature, luminosity and absolute magnitude. These sensitive parameters provide very important information about how such systems formed, how they evolved and what physical processes they undergo during their lifetimes. By studying systems such as XY Leo, it is possible to investigate many fundamental astrophysical problems in detail. High-precision TESS observations and spectral observations of XY Leo, a double-lined eclipsing binary system, have allowed us to accurately determine the parameters of its components. Simultaneous measurements for the XY Leo components are listed in Table~\ref{table:xyleo:lcsol} and plotted in the Figs~ \ref{Fig:xyleo_all_rvs} and ~ \ref{Fig:xyleo_all_lc}. 

The parameters obtained as a result of the simultaneous analysis of the radial velocity and light curves of the system, the masses and radii of the individual components, were obtained from the $M_{1,2}\sin^3\,i$ and $a\sin\,i$ relations.  The luminosities were similarly obtained from the radii and temperatures.
When calculating the physical parameters of the system's component stars, the Sun's effective temperature was taken to be 5777 K and its bolometric magnitude 4.755. The masses and radii of the hot and cold components of XY Leo~A were obtained as 0.629\,M$_{\odot}$, 0.865\,M$_{\odot}$, 0.739\,R$_{\odot}$, 0.855\,R$_{\odot}$, respectively. We have used bolometric corrections (BC1, BC2) from the tables of \cite{Flower1996ApJ...469..355F}. We then determined the absolute visual magnitude of each component. Using the absolute visual magnitude and the adopted reddening, we calculated the distance of the system to be 68 pc. This value is consistent with the Gaia parallax distance of $66.7 \pm 1.6$ pc derived from $\pi = 14.99 \pm 0.35$ mas. The component stars of XY Leo A and the basic parameters obtained for the system are listed in Table~\ref{tab:XYLeo:PhyPar} together with their errors. The errors in the table are calculated from the uncertainties of the parameters (e.g., K$_{1,2}$, a$_{1,2}$, r$_{1,2}$, P, i, T$_2$) in the analysis of the radial velocity and light curves.

The catalogue of \cite{Duflot1995A&AS..114..269D} lists the spectral type of the contact binary XY Leo~A as K0n+K0, while \cite{Barden1987ApJ...317..333B} showed that the visual companion XY Leo~B is itself a BY Dra-type binary composed of late-K to mid-M dwarf stars.
In the TESS light curves we obtained, no eclipses other than the eclipse of the binary system XY Leo A were detected. However, from the light curve analyses, the light contribution of the third body was found to be 13 per cent (Table~\ref{table:xyleo:lcsol}). Because the orbital inclination angle of XY Leo~B is unknown, we can only make an approximation for the masses of the component stars using the spectral type \citep{Pecaut2013ApJS..208....9P}. Consequently, we obtained the masses of the component stars of XY Leo~B to be 0.50\,$\rm{M_\odot}$ and 0.37\,$\rm{M_\odot}$.
Similarly, from the mass-radius-temperature relation \citep{Rappaport2017MNRAS.467.2160R} of dwarf main-sequence stars, we calculated the radii of the component stars to be 0.46 $\rm{R_\odot}$ and 0.34 $\rm{R_\odot}$ and the surface effective temperatures to be 3610~K and 3450~K. The luminosities are 0.031 $\rm{L_\odot}$ and 0.015 $\rm{L_\odot}$. With these results, the orbital inclination angle of XY Leo B was determined to be between 30 and 34$^\circ$. Finally, we obtained the mass of the possible star in the outermost orbit with an orbital period of 23\, yr depending on the orbital inclination of the outer orbit (see Table~\ref{tab:xyleo:OC}). This depends on orbital inclination, but the minimum mass that the component star could have is 0.54\,$\rm{M_\odot}$ ($i_C$=90$^\circ$). It would be 2.07\,$\rm{M_\odot}$ and 0.69\,$\rm{M_\odot}$ for $i_C$ of 30$^\circ$ and 60$^\circ$, respectively. 
The total masses inferred from LITE signals, assuming orbital inclination angles, are shown in blue (XY Leo B1+B2) and red (the speculative LITE2 component) in Fig.~\ref{Fig:xyleo:ocmass34}. These may arise from magnetic activity rather than a dynamical companion. We refrain from referring to the latter as XY Leo~C owing to the lack of dynamical confirmation.

\begin{table}
\begin{center}
\caption{Astrophysical parameters of XY Leo-A. The standard errors $\sigma$ are given in parentheses in the last digit quoted.}
\label{tab:XYLeo:PhyPar}
\begin{tabular}{lll}
\hline
Parameter                                     & A1  		  &  A2\\
\hline
Mass $M/\rm{M_\odot}$                & 0.629(9)  & 0.865(12)     \\
Radius $R/\rm{R_\odot}$              & 0.739 (7)   & 0.855(8)     \\
Temperature $T_{\rm eff}/{\rm K}$    & $4\,850 (170)$    	  & $4\,578 (160)$   \\
Luminosity  $L/\rm{L_\odot}$         & 0.271(40)       & 0.288(50)     \\
Surface gravity $\log_{10}(g/\rm{cm\,s^{-2}})$ & 4.50(11)    & 4.51(13)     \\
Bolometric magnitude $M_B$          & 6\fm15      	  & 6\fm08     \\
Absolute magnitude  $M_V$                     & 6\fm55     	  & 6\fm65      \\
Semi-major axis $a/\rm{R_{_\odot}}$  &~~~2.078(10) &		\\
Distance  $d/{\rm parsec}$                             &~~~68(3)    &      	\\
\hline
\end{tabular}
\end{center}
\end{table}

\section{Physical Interpretation of the Long-Term O--C Modulations}\label{sec:Physical_OC}

To understand the physical origin of the observed O--C variations in XY~Leo~A, we analyzed the system in a hierarchical framework, combining LITE interpretations and magnetic activity modeling based on the Applegate mechanism. The observed O--C diagram exhibits three distinct modulation features: (i) a parabolic trend due to mass transfer within the inner contact binary, (ii) a well-established LITE signal with a period of $\sim$20 years caused by the gravitational influence of the outer close binary XY~Leo~B, and (iii) additional long-term residuals that include a $\sim$23–24 year sinusoidal modulation and a lower-amplitude $\sim$14-year quasi-cyclic variation.

The first and largest modulation is attributed to XY~Leo~B, whose third-body LITE signal is clearly seen in the top panel of Figure~3. The best-fit orbital period of $P_{\rm LITE1} \approx 20$ years and derived mass function indicate the presence of a close detached binary orbiting the contact pair. The residuals after subtracting this component reveal a second periodic variation with $P_{\rm LITE2} \approx 23$ years and a semi-amplitude of $\sim$13.5 minutes.

Assuming this modulation arises from an additional component in Keplerian orbit, the projected semimajor axis of the inner binary's barycenter is:
\[
a_{3} \sin i = A_{\rm LITE2} \times c \approx 0.0094 \, \mathrm{d} \times 173.145 \approx 1.63\, \mathrm{AU},
\]
where $A_{\rm LITE2}$ is the semi-amplitude in days and $c$ is the speed of light in AU/day. At a distance of 68 pc, this corresponds to a projected angular separation of $\sim$24 milliarcseconds, marginally resolvable with current high-resolution instrumentation. The corresponding radial velocity amplitude is:
\[
K \approx \frac{2\pi a_{3} \sin i}{P \sqrt{1 - e^2}} \approx \frac{2\pi \times 1.63}{22.8 \times 0.98} \approx 3.6\, \mathrm{km\,s^{-1}},
\]
which could, in principle, be confirmed through long-term RV monitoring. However, the closeness of the outer periods ($P_3 \approx 20$ yr and $P_4 \approx 23$ yr) results in a period ratio of $\sim$1.15, violating the hierarchical stability threshold \citep{Mardling2001, Tokovinin2008}. This implies that the addition of a fourth massive body with such a configuration is dynamically unstable unless a finely tuned non-coplanar or resonant architecture is assumed.

Given this instability, we explored an alternative interpretation: that the 23-year signal may originate from magnetic cycles within the contact binary, as proposed by \citet{Applegate1992} and further developed by \citet{Lanza1998, Lanza2002}. Adopting system parameters from our LC+RV models and assuming the cooler, more convective component (A2) is magnetically active, we computed the physical parameters associated with the Applegate mechanism.

The required fractional period change is derived from the observed O--C semi-amplitude:
\[
\frac{\Delta P}{P} = \frac{2\pi \Delta T}{P_{\rm mod}} \approx 6.2 \times 10^{-5},
\]
where $\Delta T = 0.0094$ days is the semi-amplitude and $P_{\rm mod} = 23$ years. The corresponding quadrupole moment variation is:
\[
\frac{\Delta P}{P} = -9 \frac{\Delta Q}{a^2 \mu},
\]
where $a$ is the binary separation and $\mu = \frac{M_1 M_2}{M_1 + M_2}$ is the reduced mass. Using system values, we find $\Delta Q \sim 3.0 \times 10^{49}$ g\,cm$^2$.

The required energy to induce this variation is calculated using \citep{Applegate1992}:
\[
\Delta E \approx \frac{1}{6} M_{\rm shell} R^2 \left( \frac{\Delta \Omega}{\Omega} \right)^2,
\]
assuming a shell mass $M_{\rm shell} \sim 0.1\,M_\odot$, stellar radius $R$, and synchronous rotation. The computed values are: $\Delta E \sim 2.54 \times 10^{43}$ erg and $\Delta L/L \sim 3.55$, which far exceed realistic limits for magnetic activity. Thus, the Applegate mechanism cannot account for the 23-year modulation.

These results are summarized in Table~\ref{tab:applegate_results}, indicating that a fifth stellar component remains the more plausible explanation for LITE2, despite its dynamical challenges. The minimum mass derived from the mass function $f(m) = 0.0135 \, M_\odot$ corresponds to $m_4 \gtrsim 0.54\, M_\odot$ (assuming $i = 90^\circ$). Future N-body modeling is required to test this possibility.

\begin{table}[h!]
\centering
\caption{Applegate parameters for the 23-year modulation in XY~Leo~A.}
\label{tab:applegate_results}
\begin{tabular}{lcccc}
\hline
Parameter & Value & Uncertainty & Unit \\
\hline
$\Delta P / P$ & $6.2 \times 10^{-5}$ & $1.0 \times 10^{-5}$ & -- \\
$\Delta Q$ & $3.02 \times 10^{49}$ & $0.52 \times 10^{49}$ & g\,cm$^2$ \\
$\Delta J$ & $5.79 \times 10^{48}$ & $0.72 \times 10^{48}$ & g\,cm$^2$\,s$^{-1}$ \\
$\Delta E$ & $2.54 \times 10^{43}$ & $0.35 \times 10^{43}$ & erg \\
$\Delta L / L$ & $3.55$ & $0.47$ & -- \\
\hline
\end{tabular}
\end{table}

After removing both LITE components, the residuals show a third modulation with a period of 14.2 years and semi-amplitude $\sim$8 minutes. Repeating the Applegate analysis for this modulation gives $\Delta E \approx 5 \times 10^{40}$ erg and $\Delta L/L \approx 0.003$, both consistent with known values for magnetically active W~UMa-type binaries \citep{Berdyugina1999,Yakut2005ApJ...629.1055Y}. Thus, this final low-amplitude modulation is plausibly attributed to stellar magnetic activity.

To test whether the inclusion of an additional (fifth) stellar component is statistically justified, we performed a Bayesian Information Criterion (BIC) analysis following the formalism of \citet{KassRaftery1995}. A model incorporating two LITE terms—one corresponding to the known 20-year modulation from XY~Leo~B and the other to the observed 23-year modulation—along with a parabolic mass-transfer trend, yields a significantly lower BIC value ($\mathrm{BIC} = -4101.90$) than a simpler model including only one LITE term and mass transfer ($\mathrm{BIC} = -3806.99$). The resulting $\Delta \mathrm{BIC} \approx 295$ provides strong statistical support for the more complex model, favoring the inclusion of a second LITE term that may be associated with a gravitationally bound fifth stellar component in the system.

In conclusion, XY~Leo is best interpreted as a hierarchical stellar system comprising an inner contact binary undergoing mass transfer and orbited by an outer detached binary (XY~Leo~B), which accounts for the well-established 20-year LITE modulation. After subtracting this signal, we detect an additional 23-year periodicity whose energetics are incompatible with the Applegate mechanism, while its orbital configuration would likely be dynamically unstable if interpreted as arising from a fifth stellar component—though this possibility cannot be entirely dismissed.  Finally, a third modulation with a characteristic timescale of $14.2 \pm 0.8$~yr and a much lower amplitude is observed in the residuals, as quantified through a Lomb--Scargle analysis of the final $O$--$C$ data.
The physical parameters derived for this lowest-amplitude variation are consistent with the expectations of magnetic cycles in late-type contact binaries and are thus best explained by angular momentum redistribution in the convective envelope of the magnetically active secondary component.
Continued long-term eclipse timing and radial velocity monitoring will be essential to fully disentangle these effects and determine the true dynamical architecture of the XY~Leo system.

Taken together, both the statistical evidence from the BIC analysis ($\Delta \mathrm{BIC} \approx 295$; \citealt{KassRaftery1995}) and the energetic constraints derived from the Applegate mechanism indicate that the $\sim$23-year modulation cannot be explained by magnetic activity alone. Instead, these findings support the presence of an additional, gravitationally bound stellar component—a fifth star in the XY Leo system. Nonetheless, given the proximity of its period to that of XY Leo~B and the potential dynamical instability of such a configuration, we cannot conclusively confirm its nature. Further high-precision timing and long-baseline astrometry will be required to test this hypothesis.

\section{DISCUSSION AND CONCLUSION}
Our analysis reveals the nature of the XY Leo multiple system using period variation analysis, radial velocity analysis, and multicolour light curve modelling, based on newly obtained ground-based data, TESS observations and nearly 90 years of archival photometric and spectroscopic measurements from the literature. The stars that compose XY Leo A form a W-type contact system, and highly precise physical parameters of the component stars have been determined (Table \ref{tab:XYLeo:PhyPar}). The results of this study show that the primary star has a smaller mass and radius than the hotter component. As well as the parameters of the contact binary system in the innermost orbit, the parameters of XY Leo B—comprising a detached binary system in the mid-orbit—were also estimated.

Recently, \citet{Kocak2023PhDT} and \citet{Nasiroglu2025} analyzed the system’s light curve and orbital period behavior using eclipse timing variations (ETVs). In this study, we compiled 415 times of minima, including 24 new measurements from TESS, and performed a comprehensive O$-$C analysis. Our results reveal that the timing residuals are best modeled by the superposition of three components: (i) a parabolic trend due to secular period increase from mass transfer, (ii) a long-term sinusoidal variation with a period of $\sim$20 years, attributable to the detached binary companion XY~Leo~B, and (iii) an additional sinusoidal signal with a period of $\sim$23 years.

The 23-year signal cannot be plausibly explained by the Applegate mechanism (see Section~\ref{sec:Physical_OC}): our energetic calculations yield fractional luminosity variations ($\Delta L/L$) and required energy budgets that significantly exceed physical limits. Moreover, its proximity in period to XY~Leo~B ($P_4/P_3 \approx 1.2$) raises concerns of dynamical instability unless specific non-coplanar or eccentric configurations are invoked. While a gravitationally bound fifth body remains a possibility, its existence cannot be confirmed with current data and remains speculative.

Notably, after subtracting the parabolic trend and both LITE components (20-year and 23-year), a residual quasi-cyclic signal remains with a characteristic timescale of $14.2 \pm 0.8$~yr and a semi-amplitude of $\sim$8 minutes. This final modulation is energetically compatible with magnetic activity cycles on the secondary star in the contact binary, consistent with predictions of the Applegate mechanism and observations in other W~UMa-type systems. Our analysis of the $O\!-\!C$ diagram therefore reveals a layered combination of distinct signals: a parabolic trend due to ongoing mass transfer in the inner binary, a 20-year LITE modulation attributed to the detached companion pair XY~Leo~B, a 23-year periodic variation that is inconsistent with both Applegate's mechanism and long-term dynamical stability as a bound fifth body, and finally, a low-amplitude $\sim$14-year modulation that is best explained by magnetic activity cycles in the convective secondary.

Although XY~Leo is not currently listed in the \texttt{nss\_two\_body\_orbit} or \texttt{nss\_non\_single\_star} tables in Gaia DR3, the system shows anomalously high RUWE ($\sim$21) and astrometric excess noise ($\sim$2.6~mas), indicating that the standard single-star model is inadequate. 
While the known SB2 pair (XY~Leo~B) accounts for part of the system’s complexity, the remaining $O$--$C$ residuals cannot be fully explained within the framework of a single-star astrometric solution.
This is expected given the contact nature of XY Leo~A and the short orbital period of XY Leo~B, both of which are challenging for Gaia’s single-star solution framework. Future astrometric solutions from Gaia DR4 or high-angular-resolution interferometry (e.g., VLTI, CHARA) may provide crucial constraints on additional companions.

Orbital instabilities in binary systems play a crucial role in their dynamical evolution. In particular, the presence of an outer companion orbiting a binary—alongside mass transfer—can give rise to complex gravitational interactions. One of the primary mechanisms driving such instabilities is the von Zeipel–Kozai–Lidov (ZKL) effect, which can induce oscillations in eccentricity and inclination, thereby modulating mass-transfer rates and Roche lobe filling over time. Given the age of XY Leo inferred from our evolutionary models, it is plausible that the ZKL effect influenced its early development and may still play a role in its secular evolution. However, its current influence on any additional long-period modulations remains uncertain.

Beyond the ZKL mechanism, the existence of two distinct long-term period variations (approximately 20 and 23 years) raises important dynamical questions. If both signals were due to orbiting companions, their close period ratio (about 1.2) would imply a configuration prone to dynamical instability unless the system satisfies specific hierarchical conditions. Such arrangements often lead to significant gravitational interactions, potentially resulting in chaotic behaviour or orbital reconfiguration over relatively short timescales \citep{Chambers1996Icar..119..261C,Laskar1996CeMDA..64..115L}. Stability might still be maintained if the source of the 23-year signal orbit at a significantly different inclination or distance from XY Leo B. Our estimated system parameters remain broadly consistent with empirical stability criteria for hierarchical multiple-star systems \citep{Tokovinin2008MNRAS.389..925T,Wang2018ApJS..235...29W}.
A more comprehensive dynamical investigation, including detailed $N$-body simulations \citep{Aarseth2003gnbs.book.....A}, is necessary to test the long-term stability of any such configurations over astrophysical timescales. Strong mutual perturbations or interactions could undermine a purely dynamical interpretation of the observed variations and instead support the idea that at least part of the modulation arises from stellar activity.

In studies on the formation and evolution of close binary stars, the transition from a detached binary system to a semi-detached and ultimately to a contact system—under the influence of processes such as non-conservative mass loss, mass transfer, angular momentum loss, the presence of a third body or differential nuclear evolution—has been extensively discussed by \citet{Yakut2005ApJ...629.1055Y} and \citet{Eggleton2017MNRAS.468.3533E}. In particular, angular momentum evolution driven by a third body through the ZKL mechanism has been addressed in detail by \citet{Naoz2016ARA&A..54..441N}, \citet{Eggleton2001ApJ...562.1012E}, and \citet{Toonen2016ComAC...3....6T}. Besides these mechanisms, system parameters such as total mass, mass ratio, and orbital periods are also key to understanding evolutionary pathways. For instance, when the mass ratio is very small ($q<0.2$), spin angular momentum can significantly influence the evolution (e.g. V1191 Cyg, CU Tau, TV Mus, and AH Cnc; \citealt{Qian2005AJ....130..224Q,Ulas2012NewA...17...46U,Yang2015AJ....150...69Y,Kocak2020CoSka..50..508K,Kocak2023PhDT}). In contrast, systems with larger mass ratios ($q>0.7$) are less affected by this.

The XY Leo system, which consists of two well-characterized binary subsystems—one contact and one detached—and exhibits additional long-term period variations, provides an excellent testbed for studying such evolutionary effects within a single system. To investigate these effects comprehensively, accurate physical and atmospheric parameters of the components, along with the properties of the three known orbits, are essential. In this study, we obtained these quantities using all available observational data. The results are summarised in Tables~2 to 5, and the model fits are shown in Figs.\ref{Fig:xyleo_all_rvs} to \ref{Fig:xyleo_oc}. The distance to the system is estimated as 68 pc, in excellent agreement with the Gaia astrometric distance of $66.7 \pm 1.6$ pc derived from the Gaia parallax $\pi = 14.99 \pm 0.35$ mas. Despite the uncertain nature of the outer modulation, XY Leo remains a valuable astrophysical laboratory for studying the evolution of close binaries, owing to its proximity and the detailed characterization of its inner subsystems.

\begin{figure}
\centering
\includegraphics[scale=0.92]{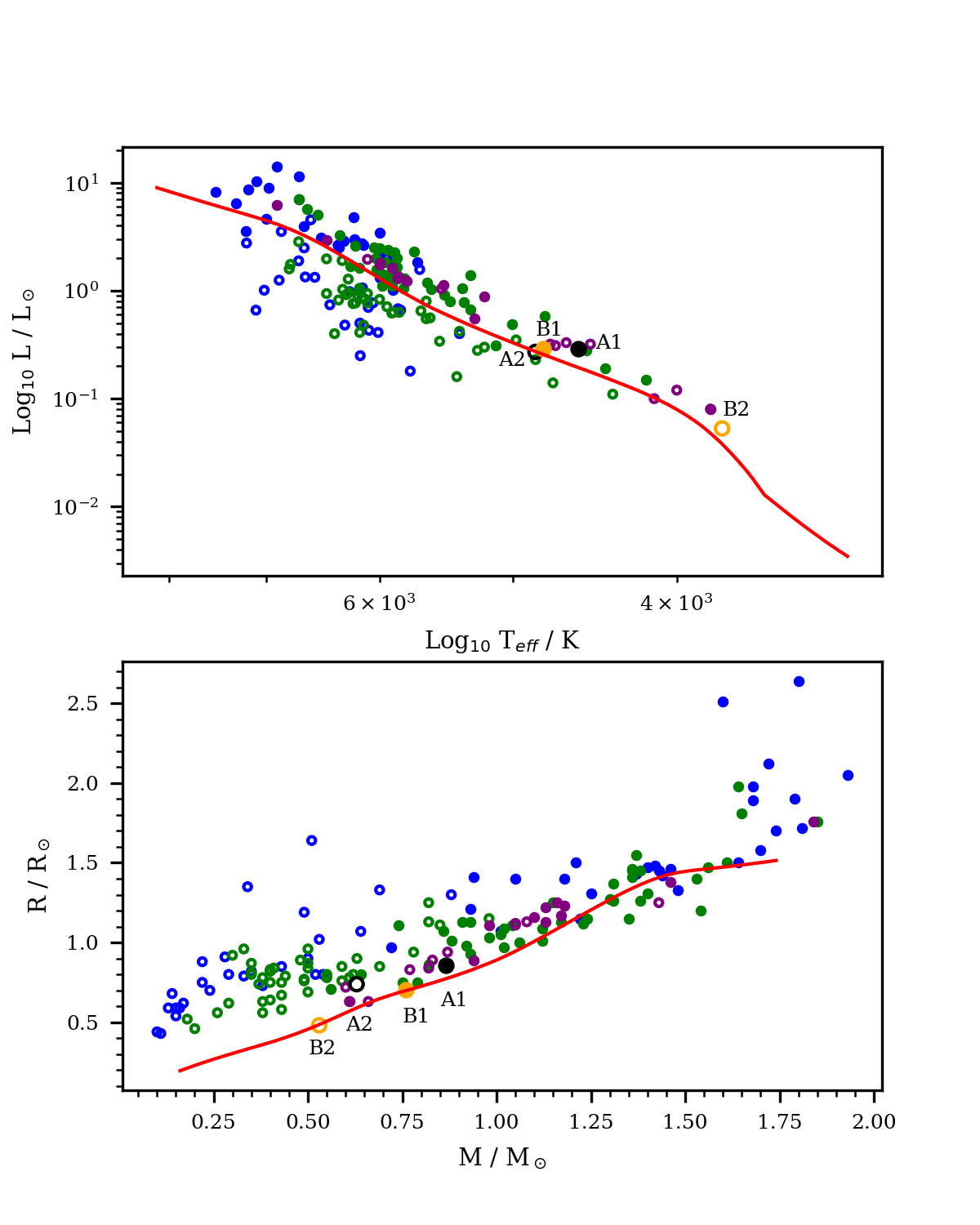}
\caption{H--R and Mass–Radius diagrams of the four confirmed component stars A1 (black filled circle), A2 (black circle), B1 (orange filled circle), and B2 (orange circle) of the XY Leo multiple system. Observed quantities for other well-studied W-type (blue), A-type (green) contact binaries, and detached systems (purple) are also shown for comparison, along with the zero-age main sequence (red lines) \citep{Yakut2005ApJ...629.1055Y,Pols1995MNRAS.274..964P}. Filled circles indicate primary components; open circles indicate secondary components.}
\label{Fig:xyleo:hr}
\end{figure}

We show the obtained physical parameters in the H–R and mass–radius diagrams, comparing them with well-characterised close binary systems and close detached binaries. The comparison systems shown in Fig.~\ref{Fig:xyleo:hr} are taken from \citet{Yakut2005ApJ...629.1055Y}. In this figure, the four securely identified components of the XY Leo system (A1, A2, B1, and B2) are plotted. The parameters we derived for these stars are in good agreement with similar systems in the literature. The proximity and detailed characterisation of the XY Leo subsystems make it a valuable target for studying the structure and evolution of close binaries.

Here we have determined the physical parameters of the XY Leo A contact binary system with high precision. To investigate its evolutionary history, we used the \textsc{EV} code \citep{eggleton2002}, a variant of the Cambridge stellar evolution models, along with its binary-star adaptation \textsc{TWIN} \citep{Yakut2005ApJ...629.1055Y}. The \textsc{TWIN} version allows the simultaneous modelling of both components of a binary system and includes non-conservative processes that affect the stellar structure and orbit. In particular, the code incorporates the effects of magnetic dynamo activity, and mass loss is assumed to carry away angular momentum through magnetic braking over the course of the system's evolution.

\begin{figure}
\centering
\includegraphics[scale=0.8]{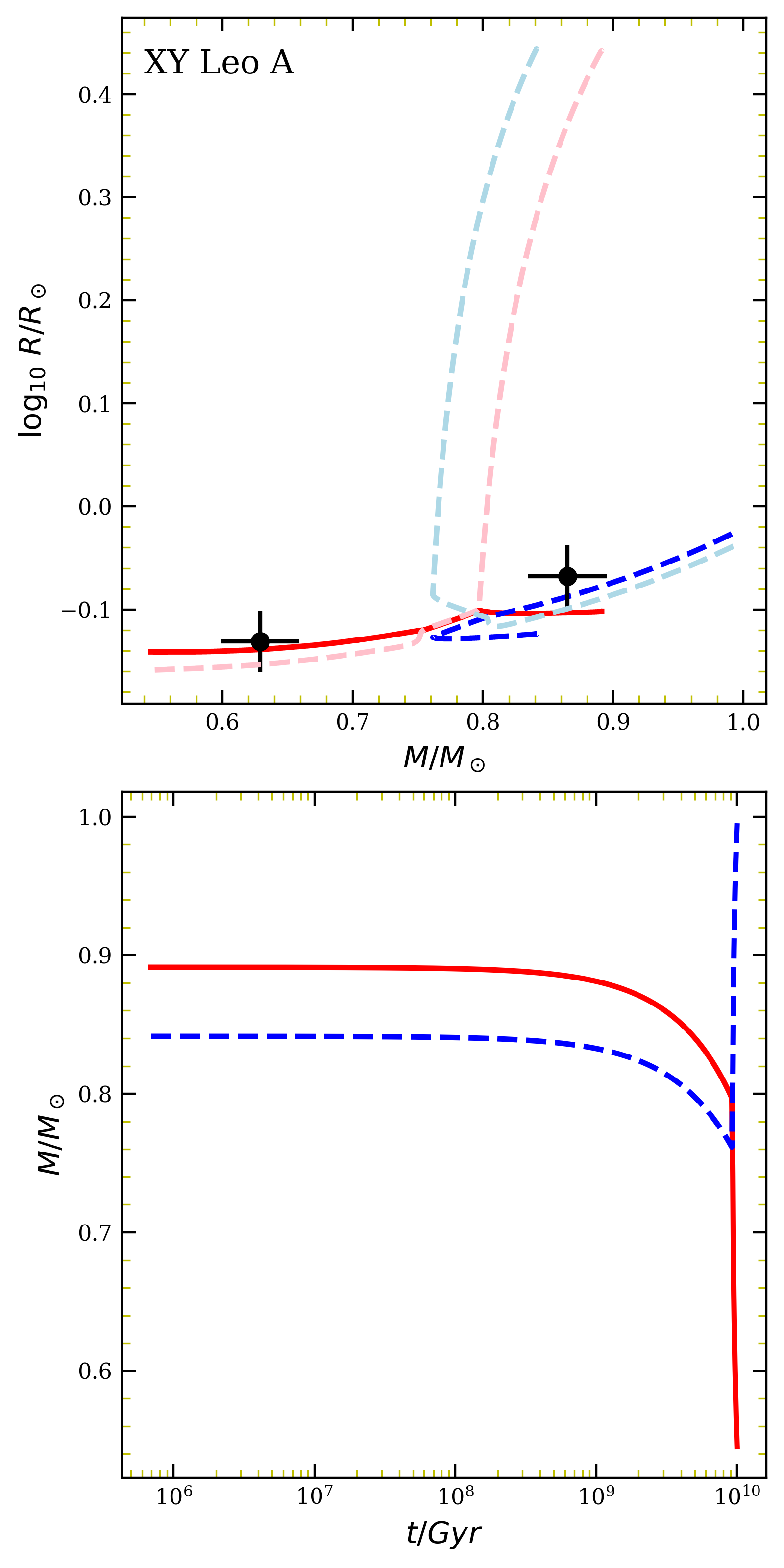}
\caption{Mass-radius (upper panel) and age-mass (bottom panel) diagrams of the XY Leo multiple system's component stars A1 and A2. Observations are shown with black points. The red and blue lines indicate the evolution tracks for components A1 and A2. The dashed pink and dashed light blue colours show the Roche lobe radii of the components. See text for details.}
\label{Fig:xyleo:ev}
\end{figure}

In the case of non-conservative evolution, identifying the initial conditions that can give rise to the present configuration of a binary system is a complex task \citep[see][]{Eggleton2017MNRAS.468.3533E}. It is often difficult to determine which component was initially more massive, particularly when significant mass transfer and angular momentum loss have occurred. However, it can generally be assumed that both components had larger initial masses than those currently observed. After exploring a range of initial mass and period configurations to model XY Leo A, we found that the best match to the present-day system parameters is obtained with initial masses of 0.94\,M$_\odot$ and 0.81\,M$_\odot$, and an initial orbital period of 0.53\,days. Angular momentum loss via wind-driven mass loss and magnetic braking was included in the models and plays a significant role in the evolution toward contact.

The modelling of non-conservative processes with the \textsc{TWIN} code shows that the system evolves to the observed masses and radii in approximately $9.5$\,Gyr. The model results are presented in Fig.~\ref{Fig:xyleo:ev}, where the red solid line traces the evolution of the initially more massive component, and the blue dashed line represents the initially less massive star. The dashed pink and light blue lines correspond to the Roche lobe radii of each component. When a star's radius intersects its Roche lobe radius, it begins to fill its Roche lobe.
According to the model, the initially more massive component fills its Roche lobe at around 9.14 Gyr, marking the onset of a semi-detached phase. At approximately 9.3 Gyr, the second component also fills its Roche lobe, and the system transitions to a contact configuration. Evolution proceeds thereafter in contact, and the present-day configuration corresponds to an age of roughly 9.5 Gyr. It is also noteworthy that the initially more massive star remains the more massive component at the present time, indicating that mass transfer has not significantly inverted the mass ratio.

Future numerical simulations and additional observational constraints are essential to assess the long-term dynamical stability of the XY Leo system. In particular, high-resolution spectroscopic monitoring is required to test whether the source of the observed long-term modulation represents a gravitationally bound companion or instead arises from magnetic activity-induced period variations. Given its proximity and well-characterised hierarchical structure, XY Leo serves as a valuable benchmark system to study the evolution of close binaries within multiple-star configurations. Continued photometric monitoring, combined with detailed dynamical modelling, will help refine its evolutionary history and clarify the nature of the outer modulation without invoking speculative additional components.

\begin{acknowledgments}
We thank the anonymous referee for the careful reading of the manuscript and for the helpful suggestions and corrections provided, which have contributed to improving the clarity of the paper.
DK thanks the Scientific and Technological Research Council of Türkiye  (T\"UB\.ITAK-2219) for her scholarship and CAT and KY thank Churchill College for their fellowships. 
Part of the data used in this study were obtained within the scope of the project numbered 18AT60-1301 conducted using the T60 telescope at the TUG (TÜBİTAK National Observatory, Antalya) site under the Türkiye National Observatories, have been utilized. The numerical calculations reported in this paper were partially performed at T\"UB\.ITAK ULAKB\.IM, High Performance and Grid Computing Center (TRUBA resources).
\end{acknowledgments}





%
\facilities{NASA Transiting Exoplanet Survey Satellite (TESS) \dataset[10.17909/fwdt-2x66]{TESS observations of XY Leo.}}

\software{astropy \citep{2013A&A...558A..33A,2018AJ....156..123A,2022ApJ...935..167A},  AstroImageJ \citep{Collins2017}, IRAF \citep{1986SPIE..627..733T}}



\bibliography{xyleo}{}
\bibliographystyle{aasjournalv7}



\end{document}